\documentclass[graybox]{svmult}

\usepackage{cite}       

\usepackage{mathptmx}       
\usepackage{helvet}         
\usepackage{courier}        
\usepackage{type1cm}        
\usepackage{makeidx}         
\usepackage{graphicx}        
\usepackage{multicol}        
\usepackage[bottom]{footmisc}

\makeindex             

\begin{document}

\title*{Sequential decoupling of negative-energy states in Douglas--Kroll--Hess theory}
\titlerunning{Douglas--Kroll--Hess theory}
\author{Markus Reiher}
\authorrunning{M. Reiher}
\institute{M. Reiher, ETH Z\"urich, Laboratorium f\"ur Physikalische Chemie, Vladimir-Prelog-Weg 2, CH-8093 Z\"urich, Switzerland, \email{markus.reiher@phys.chem.ethz.ch}}
%
%
\maketitle


\abstract*{Here, we review the historical development, current status, and prospects of Douglas--Kroll--Hess theory as a quantum chemical 
relativistic electrons-only theory.}

\section{Introduction}
In this conceptual review, we describe the development of Douglas--Kroll--Hess (DKH) theory. While we discuss the essential concepts of this theory 
explicitly, we refer the reader for further technical details to recent reviews \cite{reih06b,reih06,reih12,naka12} (for a comprehensive background
of relativistic quantum chemistry see the monograph in Ref.\ \cite{reih15}).

The symmetric occurrence of positive- and negative-energy continua in the spectrum of the
field-free Dirac Hamiltonian is the basis for states describing electrons and anti-electrons (positrons)
in quantum electrodynamics (QED). The sophisticated framework of QED is neither feasible nor necessary for a theoretical description of matter at the
molecular level. In fact, the quantization of the radiation field is hardly needed in molecular science. In a first-quantized theory, however, in which all electromagnetic interactions are
described in terms of classical fields (electromagnetic scalar and vector potentials) rather than as being transmitted by photons 
as in second-quantized QED, the negative-energy continuum creates pathologies such as variational collapse
and continuum dissolution because of the resulting boundlessness of the Dirac Hamiltonian \cite{reih15}. 

These pathologies make Dirac's relativistic theory of the electron a difficult basis for
standard numerical solution methods. Still, all technical issues can be solved in an orbital-based approach \cite{reih03,saue03,elia10b} --- mostly by
respecting the structure of the one-electron Hamiltonian (kinetic balance) 
and the proper exponentially decreasing long-range behavior of the electronic bound states when representing the one-electron positive-energy states on 
a grid or in terms of basis functions. Clearly, the negative-energy one-particle states, which are obtained whenever a one-electron operator containing the 
Dirac Hamiltonian --- such as the four-dimensional Fock operator of four-component methods ('four-component' Fock operator) --- is diagonalized, are to be omitted from the construction
of the density matrix required for the calculation of the electron--electron potential and interaction energy.

In low-energy physics and therefore also in chemistry, electron--positron pair
creation processes are energetically not accessible and therefore negative-energy states are usually not required for a sufficiently accurate quantum-mechanical
description of molecular matter. For this reason, approaches have been searched for that produce
relativistic (no-pair, electrons-only) Hamiltonians with an energy spectrum that resembles the positive-energy states only.
We should note that these states feature positive energies even for bound electronic states as the zero-energy reference is not the state, in which
all particles are found at rest and at infinitely large distance from each other. In this situation, which marks the non-relativistic zero-energy reference,
each particle still possesses a rest energy defined by the mass observed for the particle when it is at rest. 
The rest energy of an electron is given by the product of its rest mass and the square of the speed of light, $mc^2$,
and it is so large that even bound electronic states will feature a positive energy when the rest energy is added to them
(and if the attractive potential is not too strong as is the case for all atomic nuclei known).

An elegant option of removing the coupling to the charge-conjugated negative-energy states is the application of a unitary transformation $U$ that 
block-diagonalizes the Dirac Hamiltonian $h_D$ in
such a way that two decoupled operator blocks, $h_+$ and $h_-$, emerge 
\begin{equation}
\label{trafo}
h_{bd}=Uh_DU^\dagger=\left(\begin{array}{cc}h_+&{\bf 0}_2\\{\bf 0}_2&h_-\end{array}\right)
\end{equation}
where ${\bf 0}_2$ denotes a two-dimensional null matrix entering the off-diagonal blocks of $h_{bd}$. 
$h_+$ and $h_-$ then account for the positive- and negative-energy states separately.
Both,  $h_+$ and $h_-$, are two-dimensional one-electron Hamiltonians. The 2$\times$2 super-structure of $h_D$,
\begin{equation}
h_D=k_{\mathcal{O}}+r_{\mathcal{E}}+v
\end{equation}
with
\begin{equation}
k_{\mathcal{O}} = \left(\begin{array}{cc}{\bf 0}_2& c \,{\bf \sigma}\!\cdot\! {\bf p}\\
     c \,{\bf \sigma}\!\cdot\! {\bf p} &{\bf 0}_2\end{array}\right)
\end{equation}
and
\begin{equation}
r_{\mathcal{E}} = \left(\begin{array}{cc}  mc^2   &{\bf 0}_2 \\ {\bf 0}_2&-mc^2\end{array}\right)
\end{equation}
is preserved by the transformation in Eq.\ (\ref{trafo}). 
In the standard representation, the kinetic energy operator $k_{\mathcal{O}}$ is off-diagonal (odd, '${\mathcal{O}}$'), while an external electrostatic potential $v_{\mathcal{E}}$ in $v=v_{\mathcal{E}}+v_{\mathcal{O}}$ is 
block-diagonal (even, '${\mathcal{E}}$') and vector-potential contributions $v_{\mathcal{O}}$ to $v$ are odd. Also, the rest energy $r_{\mathcal{E}}$ is an even operator; hence, the subscript.

\section{Formal Exact-Decoupling}
In the mid 1980s, a formal expression for the block-diagonalizing unitary transformation $U$ was derived \cite{heul86},
\begin{equation}   
\label{eq4-38}
U  =  \left( \begin{array}{cc}
\big(1+X^{\dagger}X\big)^{-1/2} \;&
\big(1+X^{\dagger}X\big)^{-1/2}\,X^{\dagger} \\[4mm] -{\rm e}^{{\rm
i}\varphi}\big(1+XX^{\dagger}\big)^{-1/2}X \; & {\rm e}^{{\rm
i}\varphi}\big(1+XX^{\dagger}\big)^{-1/2} \end{array} \right) 
\end{equation}
as a function of the $X$-operator, which relates the large 'L' and small 'S' components,
\begin{equation}
\psi^S=X\psi^L, 
\end{equation}
of the 4-spinor, $\psi=(\psi^L,\psi^S)$. 
In their original work \cite{heul86}, Heully and co-workers
chose $\varphi \!=\! \pi$ for the relative phase $\varphi$.

An expression for $X$ depending on the energy eigenvalue $\epsilon$ can be easily derived from the Dirac equation,
\begin{equation}  
\label{eq4-2}
X = \big(\epsilon-v+2mc^{2}\big)^{-1} c \,{\bf \sigma}\!\cdot\! {\bf p} ,
\end{equation}
where $v$ is the potential energy operator, $m$ the rest mass of the electron, $c$ the speed of light, $\bf{p}$ the momentum operator,
and $\bf{\sigma}$ the 3-vector of Pauli spin matrices.

As the energy eigenvalue is the sought-for solution {\it after} applying the unitary transformation, the energy-dependent $X$-operator is not very useful. In fact,
it was possible \cite{heul86} to derive an equation for the determination of $X$ that does not depend on the energy eigenvalue,
\begin{equation} 
\label{eq4-4}
X  =  \frac{1}{2m_ec^{2}} \, \Big\{ c \,{\bf \sigma}\!\cdot\! 
{\bf p} \, - \, [X,V] \, - \, X c\,{\bf\sigma}\!\cdot\! {\bf p} X
\Big\} 
\end{equation}
However, the solution of this 
equation for $X$ was considered to be as complicated as the solution of the Dirac equation itself. However, it was not before the dawn of the new millenium that such an equation 
was solved by numerical means as a true option for exact decoupling \cite{bary02}. It was this paper by Barysz and Sadlej that introduced the first infinite-order two-component
(IOTC) method and that intiated the intense development
of exact-decoupling methods in the first decade of the 21st century.
We shall later discuss some of its ingredients in more detail (see section \ref{other}).

\section{Foldy--Wouthuysen Transformations}
Foldy and Wouthuysen were the first to find a block-diagonalizing unitary transformation in closed-form \cite{fold50}, but only for the free-particle (field-free) Dirac Hamiltonian,
for which $v=0$. Unfortunately, such a closed-form solution is not known for the general many-electron case in an external field of atomic nuclei for two reasons. (i) Already 
for a single electron in the presence of an external electrostatic potential $v_{\mathcal{E}}$, no closed-form expression for a unitary transformation can be constructed 
\cite{reih04}. (ii) Vector potentials, such as those emerging from external magnetic fields or the magnetic interaction of two electrons, as well as contributions from 
exchange integrals (in Hartree--Fock-type theories) to the off-diagonal super-block of the Dirac Hamiltonian add additional complexity to
the problem \cite{lube08}. A solution was already suggested by Foldy and Wouthuysen \cite{fold50}. A sequence of unitary transformations can be used to 
suppress the off-diagonal blocks order by order in terms of an expansion parameter. Strictly speaking, exact decoupling is then obtained only after an
infinite number of such transformations.

In physics, expansions of relativistic Hamiltonians are usually carried out order by order with respect to the inverse speed of light $1/c$, and so was the expansion
of Foldy and Wouthuysen. A $1/c$-expansion is obviously advantageous as it easily allows us to derive the non-relativistic limit for $c\rightarrow\infty$, which
should match the Schr\"odinger Hamiltonian. This holds for the Foldy--Wouthuysen $1/c$-expansion.
Already from the free-particle Foldy--Wouthuysen transformation, the one-electron Pauli Hamiltonian emerges 
at second order in $1/c$, which provides the lowest-order one-electron mass-velocity, Darwin, and spin--orbit corrections to the Schr\"odinger Hamiltonian. 
If the free-particle transformation is applied to the four-component many-electron Hamiltonian that
includes Coulomb and Breit interactions of the electrons, the Breit--Pauli Hamiltonian will result as zeroth- to second-order terms
\cite{chra53,chra53b,bark55}. 

The Pauli Hamiltonian is known to be useful in
a perturbation-theory context, but produces difficulties when applied in a variational approach. We have argued \cite{reih04} that all $1/c$ expansions --- also the
one produced by a sequence of unitary transformations as proposed by Foldy and Wouthuysen --- will fail in a variational context as the true expansion parameter is actually
the momentum divided by $mc$, which should be smaller than one for a Taylor expansion of the relativistic energy--momentum relation to converge. This, however,
cannot be guaranteed as  
can be understood in terms of formal and physical reasons.
On the one hand, high-momentum eigenfunctions cannot be excluded from a complete-basis-set representation of the $1/c$-expanded Dirac Hamiltonian and so $p/(mc)<1$
cannot be guaranteed for all such basis functions.
On the other hand, an electron may acquire high momentum in the close vicinity of heavy nuclei that may produce a case in which $p/(mc)>1$.

\section{Douglas--Kroll Transformations}
An alternative is the expansion of the Hamiltonian in terms of the potential $v$ as a formal expansion parameter. It often goes without saying explicitly that $v_{\mathcal{E}}$ and not the
full $v$ is chosen as an expansion parameter. This has dramatic consequences as $v_{\mathcal{E}}$ is the even part of $v$ and its odd complement, $v_{\mathcal{O}}$, containing vector-potential 
and exchange contributions 
is not considered in the transformation procedure. 
The inclusion of vector potentials amounts to additional difficulties, which are not discussed in this chapter. Instead, we may refer the reader to
Ref.\ \cite{lube08} and references cited therein for a detailed discussion of $v_{\mathcal{O}}$ in the context of transformation techniques.
Moreover, by default $v_{\mathcal{E}}$ contains only the external electrostatic potential of the nuclei, because the block-diagonal contribution from the electron--electron interaction
is not easy to evaluate (it depends on the ansatz for the wavefunction approximation and requires an iterative, self-consistent determination). Hence, 
a well-defined
standard choice is $v_{\mathcal{E}}=-\sum_A Z_A/r_A$ as the sole contribution ($Z_A$ is the nuclear charge number of nucleus $A$ and $r_A$ is the length of the difference vector of its positions 
to that of an electron, all in Hartree atomic units).

Expansions in terms of the inverse speed of light and of the potential have been considered by Erikson and co-workers at around 1960 \cite{erik58,erik58b,erik60,erik61}, which apparently 
has never been recognized in quantum chemistry. In 1974, an expansion of the Dirac Hamiltonian in terms of the external electrostatic
potential was proposed by Douglas and Kroll in the appendix of their paper \cite{doug74}. In the mid 1980s, Hess discovered this appendix  \cite{hess85,hess86,jans89}
and combined it with a smart computational protocol to evaluate the momentum-space expressions of the electrons-only Hamiltonian 
in a basis of position-space one-electron functions (such as Gaussian functions used in almost all
molecular quantum chemistry computer programs). 

The essential insight by Hess \cite{hess85} was that for practical applications eigenvalues of the squared momentum operator are required, which
are known in a position-space basis that diagonalizes the matrix representation of the $p^2$-operator. Since the non-relativistic kinetic energy operator $-\hbar^2\Delta/(2m)$ contains
the squared momentum operator, $p^2=-\hbar^2\Delta$, an operator in a position-space basis can be transformed into one in the $p^2$-basis by transforming it with the eigenvectors of
the kinetic energy matrix scaled by $2m$. As a fortunate consequence, an explicit momentum-space respresentation of all operators is not
necessary (and would have been a significant obstacle for an implementation of the Douglas--Kroll--Hess approach in a standard quantum-chemistry
program designed for molecular applications).
It was explicitly shown by Liu and co-workers that exploiting eigenvectors of the $p^2$-operator corresponds to the choice of a 
kinetically balanced basis set and that the matrix representation of Hess' (DKH) Hamiltonian can thus be derived from the matrix form of the Dirac Hamiltonian in such a basis set \cite{liu-10,li--12}.

In his original work on the DKH Hamiltonian \cite{hess86}, Hess considered all terms in the transformed Hamiltonian up to second order in the external
electrostatic potential, which defines the second-order DKH Hamiltonian $h_{+}^{\rm DKH2}$. This derivation had to be slightly corrected in Ref.\ \cite{jans89}.
It was not before the year 2000 that the third-order Hamiltonian was derived and applied
in quantum chemical calculations \cite{naka00}, followed by the correct fourth- and fifth-order DKH Hamiltonians \cite{wolf02b}, and the sixth-order
one \cite{vanw04}. Note that the fourth- and fifth-order Hamiltonians in Ref.\ \cite{naka00} turned out to be not correct \cite{wolf02b}. Moreover, Ref.\ \cite{naka00} presents results only for
the third-order Hamiltonian, which, however, do not show the correct (oscillatory) convergence behavior (see below). 
An arbitrary-order and therefore exact numerical decoupling approach in terms of DKH Hamiltonians was then considered by us in 2004 \cite{reih04}.

While analytic results on the boundedness of the second-order DKH Hamiltonian could be obtained \cite{brum02,sied05,sied06}, only the first implementation 
of the arbitrary-order DKH approach \cite{reih04b} demonstrated the order-by-order convergence, and variational stability could be (numerically) investigated for high orders.
The order-by-order convergence can be understood in terms of the true rather than the formal
expansion parameter \cite{reih04}: that is the potential (expressed in terms of matrix elements in the given $p^2$-basis) divided by huge energy denominators.
However, we found an oscillatory convergence behavior\cite{reih04b}: Odd DKH orders yield energy eigenvalues that are below the Dirac reference energy, while even DKH orders
approach the Dirac reference from above. This behavior can be understood in view of the sign of the leading term in the truncation error \cite{reih04,reih04b}. 

For the sequential order-by-order decoupling of the Dirac Hamiltonian in DKH theory, the necessary first step \cite{reih04} is a free-particle Foldy--Wouthuysen transformation $U_0$ to
generate an odd operator $\mathcal{O}_1$ that is linear in the potential,
\begin{equation}
h_{1}=U_0h_DU_0^\dagger= \mathcal{E}_0 + \mathcal{E}_1 + \mathcal{O}_1,
\end{equation}
besides two four-dimensional even operators, $\mathcal{E}_0$ and $\mathcal{E}_1$
(the subscript denotes the order in the potential $v_\mathcal{E}$), 
which remain unchanged under all subsequent transformations.
They define the first-order DKH Hamiltonian,
\begin{equation}
h_+^{\rm DKH1}= \mathcal{E}_0 + \mathcal{E}_1 .
\end{equation}
For explicit expressions of the low-order even terms see Ref.\ \cite{wolf02b}.

The subsequent transformations (in principle, infinitely many of them), are chosen to eliminate the lowest-order odd term at a given step. Hence, $U_1$ is chosen such that
$\mathcal{O}_1$ is eliminated, while new odd terms of higher order emerge. Then, $U_2$ eliminates $\mathcal{O}_2$ and so forth. Fortunately, each of these unitary transformations
produces two even orders that remain unchanged by the higher-order unitary transformations. I.e., $U_1$ produces the final expression for $\mathcal{E}_2$ and $\mathcal{E}_3$, while
$U_2$ produces $\mathcal{E}_4$ and $\mathcal{E}_5$ and so on. This has been called the $(2n+1)$-rule for producing the $(2n+1)$th-order DKH Hamiltonian from $U=U_nU_{(n-1)}\dots U_0$.

The order-by-order elimination of odd operators in the Hamiltonian is achieved by choosing the parameter $W$ that parametrizes
the unitary transformation at a given step in such a way that the lowest-order odd operator of that step, to which $W$ contributes, cancels.
Many closed-form expressions for the parametrization of the unitary transformation are available. For example, Douglas and Kroll
\cite{doug74} proposed the so-called square-root parametrization 
\begin{equation}
U_i^{\rm SQR}=\sqrt{1+W_i^2}+W_i, 
\end{equation}
while Nakajima and Hirao \cite{naka00} employed the exponential parametrization, 
\begin{equation}
U_i^{\rm EXP}=\exp(W_i) ,
\end{equation}
which is known best in quantum chemistry. 

When the exponential parametrization was applied in the derivation of the low-order DKH Hamiltonians by Nakajima and Hirao \cite{naka00}, it was not clear whether
these low-order Hamiltonians are actually independent of the parametrization chosen. Since the analytic parametrizations are expanded in a Taylor series expansion 
in powers of the anti-hermitean parameter $W_i$, we set out \cite{wolf02b} to study the most general unitary transformation, in which unitarity is imposed on the
coefficients of a general power series expansion in terms of $W_i$. This ansatz covers all possible, and thus infinitely many, parametrizations of the unitary transformation.
We found \cite{wolf02b} that only up to fourth order is the DKH Hamiltonian independent of the chosen parameterization. Starting at fifth order, DKH Hamiltonians depend on the
expansion coefficients of the unitary transformation, an unfortunate effect that vanishes only at infinite order.
However, the parameter dependence of the fifth- and all higher-order DKH Hamiltonians is, for reasonable parametrizations of the unitary transformation, much smaller than the
amplitude of the oscillatory convergence with increasing DKH orders \cite{reih07b}. 

It is clear that, at infinite-order, any unitary transformation will exactly reproduce the spectrum of the Dirac Hamiltonian. However,
this does not hold for the eigenstates. Different unitary transformations produce different DKH wavefunctions and different DKH
orbitals at some given order (and also at infinite order). Only 
expectation values in the four-component theory are preserved by unitary transformations of the wavefunction and the property operator. The according transformation of the
property operator has occasionally been omitted as the error introduced --- the so-called picture-change error \cite{baer90} --- is small for valence-shell properties. 
However, it can be significant and therefore non-negligible for properties probed closed to an atomic nucleus \cite{wolf06b}. A most prominent example, in which the picture-change
error is dramatic, is the contact electron density \cite{mast08b}, which is central to calculating the M\"ossbauer isomer  shift \cite{baro08,knec11}.

For properties, the $(2n+1)$-rule does not hold and $n$ unitary transformations are required to produce an $n$th-order DKH property operator \cite{wolf06}.
A symbolic scheme for the automated derivation of arbitrary-order DKH property operators has been presented \cite{wolf06b}.

Note that the DKH expansion is {\it not} of the type that yields the Schr\"odinger Hamiltonian to lowest order (that is only achieved by considering the limiting
case of $c\to\infty$). Accordingly, one does not obtain 'relativistic corrections' in the DKH expansion as there is no non-relativistic (zeroth-order) reference.

As a final remark, we should emphasize that the derivation of any DKH Hamiltonian produces a four-dimensional operator, i.e., one that contains an approximation to $h_+$ as well as to
$h_-$ on the block-diagonal. The approximation for $h_+$ is then obtained by replacing the Dirac parameter matrix $\beta$=diag$(1,1,-1,-1)$ in all terms of the DKH expansion by the
two-dimensional unit matrix.

\section{Implementation of Douglas--Kroll Transformations}
While Hess and others derived the lowest- and low-order DKH Hamiltonians manually, it became apparent that this step-wise decoupling protocol
can be fully automated \cite{reih04,reih04b}, also for molecular properties \cite{wolf06,wolf06b}. The derivation of an arbitrary-order DKH Hamiltonian or property operator
was accomplished fully symbolically in two steps \cite{reih04b,wolf06b}. First, all terms contributing to a given order were derived on a rather abstract formal level in terms
of even and odd operators of a well-defined order in the external potential.
Then, the resolution of the identity, $({\bf\sigma}\cdot{\bf p})({\bf\sigma}\cdot{\bf p})/p^2$,
is used to break down all expressions into matrix products of known non-relativistic operator matrices 
plus two additional types of 'relativistic' matrices, namely those of the operators
${\bf p}\cdot v{\bf p}$ and ${\bf p}\cdot O{\bf p}$ (where the omission of the vector ${\bf\sigma}$ of Pauli spin matrices in
front of each momentum operator indicates the standard
spin-free one-component approximation to the DKH approach)
required for the decoupled one-electron Hamiltonian and property operator $O$, respectively.

Unfortunately, this two-step protocol produces operator expressions of increasing length
(measured by the number of matrix multiplications) with increasing order. Because of this steep scaling, DKH operators up to fourteenth order were considered in the early years \cite{reih04b,wolf06b}.
Peng and Hirao realized that the cost of the whole derivation can be significantly reduced by avoiding the second step \cite{peng09} so that DKH calculations up to
35th order were easily possible \cite{peng12,peng13}. We should note, however, that already the low-order DKH Hamiltonians, and in particular the original DKH2 one, provide an accurate
description of valence-shell properties (see Refs.\ \cite{hess00b,wolf04} for two examples). Only properties probed close to an atomic nucleus \cite{wolf06b,mast07,reih07b} such 
as contact densities \cite{mast08b} or core excitations in X-ray and UV spectroscopies require high orders.

Amazingly, the step-wise derivation of Foldy--Wouthuysen decoupling in powers of $1/c$ had already been automated in 1968 on a Telefunken TR4 computer in ALGOL60 
by deVries and Jonker \cite{devr68,jonk68}. The even and odd decomposition of the Hamiltonian was achieved by mapping this digital structure to a binary number. In this
way, the Hamiltonian could be derived 'semi-symbolically' up to 8th order in $p/(mc)$ in Ref.\ \cite{devr68} and to 10th order in $p/(mc)$ in Ref.\ \cite{jonk68}.

The DKH approach is best known in its scalar-relativistic variant, in which all spin-dependent terms are separated from the scalar ones 
(by application of Dirac's relation) and then omitted. Clearly, omitting all Pauli spin matrices from the DKH Hamiltonian eliminates the
spin--orbit coupling and a spin-averaged description emerges. The resulting scalar DKH Hamiltonian still comprises all kinematic relativistic effects (to arbitrary order in the
potential). As its eigenfunctions are scalar functions, it can be easily interfaced with any non-relativistic quantum chemistry computer program. The non-relativistic
one-electron Hamiltonian in the Fock operator is then replaced by a scalar-relativistic DKH Hamiltonian of pre-defined order.
Various corrections to improve on the standard approximations in practical applications of DKH theory have been proposed 
\cite{samz91,samz92,park94,hess96,schi98,schi98b,boet00,vanw05,auts12b}.

\section{Relation to other Exact-Decoupling Approaches\label{other}}
A hybrid approach first proposed by Jensen \cite{jens05} and then elaborated by Liu, Kutzelnigg,
Saue, Visscher, Ilia\v{s}, and co-workers \cite{kutz05,kutz06,liu-06,fila06,kutz06b,ilia07,liu-07,peng07,sikk09,liu-09} offers the
possibility to achieve decoupling by a single unitary transformation. 
It is now common to denote this approach the exact two-component approach, with acronym 'X2C'. The central idea is that one can construct
an exact unitary transformation matrix $U=U(X)$ from the eigenvectors of the four-component Fock operator through their relation to the $X$-operator
in matrix representation, ${\bf X}$, 
\begin{equation}
C_{S}^{(+)}={\bf X}C_{L}^{(+)} \quad \Rightarrow \quad {\bf X}=C_{S}^{(+)}\left(C_{L}^{(+)}\right)^{-1}
\end{equation}
where $(C_{L}^{(+)},C_{S}^{(+)})$ contain the positive-energy '($+$)' eigenvectors of the four-component Fock matrix (they are the expansion coefficients of the basis-set expansions
for the large '$L$' and small '$S$' two-spinors of the four-component molecular orbital).
Clearly, this implies that the Fock operator needs to
be diagonalized first (in a given one-electron basis set), and so the problem seems to have already been solved then (it also implies that the solution of a four-component
problem is actually feasible). However, if this calculation is done only for an approximate Fock operator, in which the electron--electron
interaction terms are neglected or approximated, then an efficient approximation to the basis-set representation of the exact unitary transformation can be obtained,
which produces an approximate two-component Fock operator to which missing potential energy terms (most importantly, the full electron--electron interaction) are
added. This procedure will produce a picture-change error for all interaction terms that were not considered in the construction of the unitary transformation. 
However, the resulting X2C Hamiltonian reproduces the original
spectrum of the (full) four-component (reference) Fock Hamiltonian well so that it can be employed in a two-component electrons-only theory.

The above-mentioned IOTC method of Barysz and Sadlej \cite{bary02} is actually an extended X2C approach that involves one additional unitary
transformation, namely the free-particle Foldy--Wouthuysen transformation. Although the latter is the essential first ingredient of DKH theory,
it is not mandatory in an X2C-type approach and therefore only increases the computational cost. In order to avoid confusion due to the rather general acronym
'IOTC', the two-step exact-decoupling approach by Barysz and Sadlej has often been called the 'BSS' approach according to the initials of the authors of 
an earlier paper\cite{bary97} to which it is related.

It is important to understand that the computational effort for two-component decoupling approximations scales with a measure of the size of the system under consideration, e.g., with the number of basis functions.
Accordingly, an efficient systematic approach to the decoupling transformation
considers its atomic composition, which can be particularly easily achieved in the case of atom-centered basis functions, to set up local decoupling approximations. An atomic
decomposition of the DKH unitary transformation has been proposed by us \cite{peng12b} and by Seino and Nakatamo \cite{sein12,sein12b}, who also developed a
geometry gradient for structure optimizations \cite{naka13}. An alternative is the atomic-decomposition of the Hamiltonian \cite{gagl98,pera04,pera05,thar08},
which, however, produces no general recipe for the transformation of off-diagonal atom--other-atom blocks in the Hamiltonian.
Our derivation of the diagonal local approximation to the unitary transformation (DLU) \cite{peng12b}
was sufficiently general to also comprise the local X2C and BSS approaches. Hence, the computational effort is reduced by a reduction of dimension of matrices subjected to
multiplication, inversion, and diagonalization operations in DKH, BSS, and X2C, rather than by a limitation of the decoupling accuracy by truncation of a series expansion as in DKH2. 

We have presented
a highly efficient implementation of the local exact-decoupling methods in the {\sc Turbomole} program package along with a detailed analysis of the performance
of the different approaches \cite{peng13}. 
Moreover, we refer the reader to a general overview \cite{peng12}, which provides a detailed
numerical comparison of different exact-decoupling approaches.

\section{Conclusions and Outlook}
The present situation in relativistic quantum chemistry is such that all potential pitfalls associated with a four-component
many-electron theory based on the Dirac one-electron Hamiltonian can be circumvented by an appropriate expansion of the molecular orbitals (spinors) into
a finite one-electron basis set in such a way that all properties of the underlying Hamiltonian are respected by this expansion (kinetic balance).
Even the dimension of the matrix representation of a four-component Fock operator can be limited to be about twice as large as the one for
a corresponding non-relativistic Schr\"odinger-based Fock Hamiltonian. Consequently, computational difficulties can no longer be hold account for
the development  of two-component methods. 
In fact, four-component relativistic calculations have become routine, but since not many groups are working
in this field, we can be grateful to the major effort of the {\sc Dirac} development team \cite{DIRAC14} that an open-source, freely available,
highly professional, multi-purpose, general, four-component molecular electronic structure program is available. 

As a consequence, we finally need to address the question whether 
two-component approaches are still of value in computational quantum chemistry or whether they will be eventually replaced by four-component methods
valid for the whole periodic table of the elements. The discussion of this question has a long history \cite{quin98a} and we shall only touch upon it from
the point of view of practical molecular electronic-structure calculations.

An important case can be made for sophisticated relativistic electron-correlation methods such as (four-com\-po\-nent) multi-configurational self-consistent-field (MCSCF) \cite{jens96,abe-06},
(four-com\-po\-nent) coupled cluster (CC) \cite{elia94b,elia94c,viss96}, and (four-component) density matrix renormalization group (DMRG) \cite{knec14}. They require a
four-index transformation which switches from one- and two-electron integrals given in the atomic-orbital basis
to those in the molecular-orbital basis, in which the second-quantized Hamiltonian, that is the basis of all {\it ab initio} electron-correlation methods, is formulated.
The transformation scales with the fifth power of the dimension of the problem and is thus particularly cumbersome for four-component methods
due to the additional basis set for the small components of the molecular spinors. The requirements in terms of computational resources can be so large that this step
may prevent one from carrying out a four-component electron-correlation calculation with an MCSCF, CC, or DMRG approach. Hence, for this step, a two-component
method is beneficial as it basically requires the same effort in the four-index transformation step as a nonrelativistic approach would require.

Clearly, the X2C approach, especially in its local version \cite{peng12b}, is most efficient for this purpose and so the next question is whether 
approximate and sequential decoupling approaches will continue to have a right to exist in computational chemistry. Clearly, DKH2 will be around for some time
as many developments have been based on this low-order Hamiltonian (and its accuracy for valence-shell properties such as vibrational frequencies, reaction energies, 
or bond lengths is undeniable). The most important technical advantage of DKH2 is the supply of basis sets for the whole periodic table of the elements that were
produced by many groups in the past two decades. 

Moreover, approximate two-component methods such as DKH2 require less computational effort than X2C
and so they may be beneficial for extensive calculation, if the calculation of the one-electron Hamiltonian is the limiting step,
which can, however, be circumvented by introducing the DLU approximation \cite{peng12b}. 

All analysis in this conceptual overview focused mostly on the Hamiltonian and thus on the energy as the target observable. For other observables or
specific electronic-structure methods, it may still be advantageous to consider a sequential decoupling protocol \cite{li--12}.

Apart from these computational considerations, we should not forget that, at its heart, DKH theory is an analytic tool for deriving an electrons-only
Hamiltonian in the no-pair approximation of first-quantized relativistic many-electron theory. As such it will persist as the unique decoupling protocol
yielding variationally stable Hamiltonians.

\section*{Acknowledgments}
This work was financially supported by the Swiss National Science Foundation SNF and by ETH Zurich.


\end{document}